\documentclass[letterpaper,twocolumn,10pt]{article}
\usepackage{usenix2019_v3}

\usepackage{subfig}
\usepackage{multirow}
\usepackage{graphicx}
\usepackage[title]{appendix}
\usepackage{color}

\graphicspath{{./figures/}}

\newcommand{\boldtitle}[1]{\vspace{5px}\noindent\textbf{#1}}

\begin{document}

\date{}

\title{\Large \bf XSS Vulnerabilities in Cloud-Application Add-Ons}

\author{
{\rm Thanh Bui} \\ Aalto University
\and
{\rm Siddharth Rao} \\ Aalto University
\and
{\rm Markku Antikainen} \\ Aalto University
\and
{\rm Tuomas Aura} \\ Aalto University
} 

\maketitle


\begin{abstract}
Cloud-application add-ons are microservices that extend the functionality of the core applications. Many application vendors have opened their APIs for third-party developers and created marketplaces for add-ons (also add-ins or apps). This is a relatively new phenomenon, and its effects on the application security have not been widely studied. It seems likely that some of the add-ons have lower code quality than the core applications themselves and, thus, may bring in security vulnerabilities. We found that many such add-ons are vulnerable to cross-site scripting (XSS). The attacker can take advantage of the document-sharing and messaging features of the cloud applications to send malicious input to them. The vulnerable add-ons then execute client-side JavaScript from the carefully crafted malicious input. In a major analysis effort, we systematically studied 300 add-ons for three popular application suites, namely Microsoft Office Online, G Suite and Shopify, and discovered a significant percentage of vulnerable add-ons in each marketplace. We present the results of this study, as well as analyze the add-on architectures to understand how the XSS vulnerabilities can be exploited and how the threat can be mitigated.
\end{abstract}


\section{Introduction}
\label{sec:introduction}

In modern web applications, user data is stored and processed mainly in the cloud, and the user interface is implemented with HTML and JavaScript on the web browser. This kind of architecture has several advantages. For example, the users do not need to install or update the applications, and sharing and synchronizing data between users and services become easier. A well-known example of a cloud application is Google Docs~\cite{googledocs}, an online document editor, which allows collaborative editing of office documents. The users naturally need to trust the cloud platforms to keep their data safe, and cloud-application developers have come a long way in securing the services.

Many of the cloud applications follow the microservice architecture where much of the functionality is implemented as independent services that are loosely coupled to the core service through APIs. The APIs can also be opened to external developers. The features implemented with these APIs are variably called \emph{add-ons}, \emph{add-ins}, or \emph{apps}; we use the word add-on in this paper. For instance, the Translate~\cite{translateaddon} add-on for Google Docs allows the user to translate text to a chosen language --- a feature that is not part of the core service. Successful cloud applications have created \emph{marketplaces} for add-ons and aim to grow an ecosystem of innovative add-on services around their core platform.

The growing add-on market, however, creates new dangers. Many add-ons are quick hacks by inexperienced developers, and the users may not be aware of the difference between the add-on and the trusted host platform. Moreover, the host-application vendors are under pressure to attract new add-on developers, which can lead to less stringent security controls for the add-ons than for the core service.

In this work, we study the security risks that arise from potential security vulnerabilities in the add-ons. In particular, we are interested in how the add-on services process untrusted user input. This is a critical issue because the emphasis on collaboration and data sharing in the cloud applications makes it easy to exploit vulnerabilities in the handling of untrusted data.


The focus of our analysis is on JavaScript code injection, popularly known as \emph{cross-site scripting (XSS)}~\cite{owaspxss}. Being a prevalent attack~\cite{mitrecve}, XSS has received much attention among web security researchers~\cite{klein2005dom,lekies201325,steffens2019don,stock2015facepalm,grossman2007xss}. However, to our best knowledge, the dangers of XSS in the context of cloud-application add-ons have not been extensively studied. In this paper, we aim to fill this gap with the following contributions:

\begin{itemize}
  \item We explain in detail how XSS attacks against cloud-application users can occur through vulnerable add-ons.

  \item We analyze the architecture designs and the security mechanisms of three popular application suites, namely Microsoft Office Online~\cite{officemarketplace}, G Suite~\cite{gsuitemarketplace}, and Shopify~\cite{shopifymarketplace}. The goal is to find what the XSS attacker can gain in each case.

  \item We evaluate how widespread the problem is with an empirical study on the add-ons from the marketplaces of the selected application suites.

  \item For defensive solutions, we discuss good practices that add-on developers can follow to secure their products. We also present the lessons that we learned from our analysis about design choices and their impact on the security of an add-on system. We hope that the lessons would be useful for any cloud-application vendors which are developing or improving their add-on systems.

\end{itemize}


The rest of the paper is structured as follows. Section~\ref{sec:background} provides necessary background information. Section~\ref{sec:xss} explains in detail how XSS can occur in vulnerable add-ons. Section~\ref{sec:architecture} describes our analysis on the add-on architectures of the selected cloud-application suites. Section~\ref{sec:EmpAnalysis} presents the results of the empirical study. Section~\ref{sec:defenses} considers defensive solutions. Section~\ref{sec:discussion} discusses the results. Finally, Section~\ref{sec:related} summarizes related work, and Section~\ref{sec:conclusion} concludes the paper.


\section{Background}
\label{sec:background}

This section explains the concepts needed in the rest of the paper, i.e.~cross-site scripting and cloud-application add-ons.


\subsection{Cross-site scripting}
\label{sec:background_xss}

Cross-site scripting (XSS)~\cite{owaspxss} is one of the most common vulnerabilities in web applications. In the XSS attack, the attacker injects malicious client-side code, typically JavaScript, to a website that does not sufficiently filter client input. When honest users access the target site, the injected code is executed in their web browsers in the same context as legitimate scripts on the same page. Thus, the injected code may gain unauthorized access to resources on the target site, bypassing user authentication and the same-origin policy (SOP)~\cite{zalewski2012tangled}.

XSS was first discussed in 2000~\cite{center2000cert}, and various variants of the attack have been discovered since then. In general, XSS attacks can be classified into four types:

\begin{enumerate}
  \item \emph{Stored (persistent) XSS}: The injected script is permanently stored in a database on the target website. It is pushed to the victim's browser when the victim accesses the stored information.

  \item \emph{Reflected (non-persistent) XSS}: The injected script is not stored on the server. Rather, the attacker tricks the victim's browser into sending the script to the server, which includes it in the immediate response page such as an error message or search results.

  \item \emph{DOM-based XSS}: The injected script never reaches the server. Instead, it is injected to the Document Object Model (DOM) of the vulnerable web page, e.g.~from a URL query string or fragment identifier, and executed directly on the client side.

  \item \emph{Persistent client-side XSS}: This is relatively new type of XSS~\cite{steffens2019don}. In the attack, the attacker injects malicious payloads into client-side storage (e.g. Web Storage, cookies) of the users that it targets. This way, if the Javascript code of the website executes the malicious data from the storage, the attack succeeds.
\end{enumerate}

XSS variants can also be classified into \emph{server-side XSS} and \emph{client-side XSS}~\cite{owaspxsstypes}. The former occurs when untrusted user data is included in an HTML response generated by the server, while the latter occurs when a JavaScript call uses untrusted user data to update the DOM of the vulnerable page in an unsafe way.

In all types of XSS, the attacker gains access to any information which the victim's browser stores or processes for the target website. Most commonly, the attacker steals cookies that enable it to impersonate the victim to the website. The attacker can also create a JavaScript key logger to record sensitive data entered by the victim, for example, passwords and credit card numbers. Moreover, the injected code can invoke HTML5 APIs, such as webcam or geolocation, although some of the APIs will only allow access if the victim has opted in to the features for the target site.

\begin{figure*}[tb]
  \subfloat[With shared workspace]{\label{fig:threat_model_colleagues}
    \includegraphics[scale=0.7]{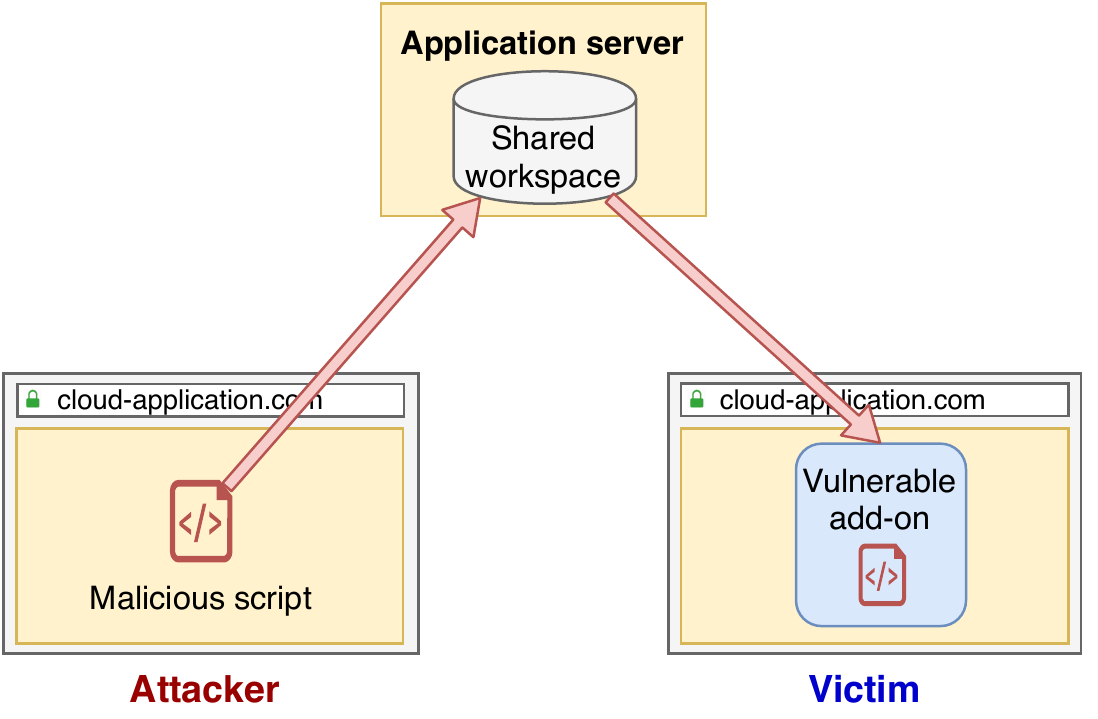}
  }
  \hspace*{\fill}
  \subfloat[With outside input]{\label{fig:threat_model_external}
    \includegraphics[scale=0.7]{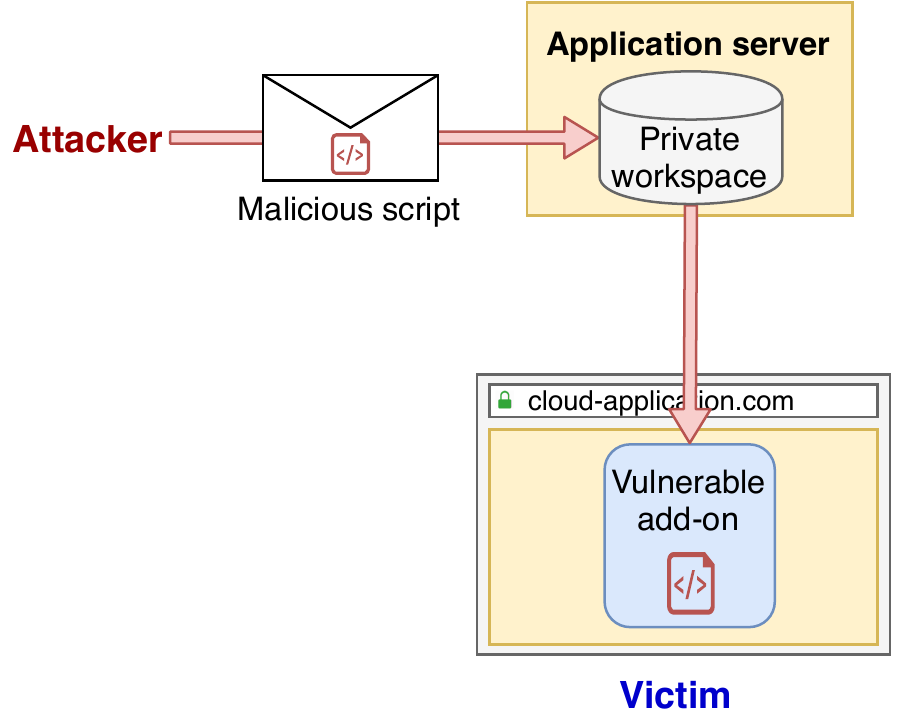}
  }
  \caption{XSS attacks with vulnerable add-ons}
  \label{fig:threat_model}
\end{figure*}


\subsection{Cloud-application add-ons}
\label{sec:background_addons}

\emph{Add-on} (also known as add-in, plugin, extension, or app) adds customized commands and features to a cloud application, called the \textit{host application}. The add-on is basically a separate web service with its own server and client components, but it has access to the user data and some core functionality of the host application through APIs defined by the application.

In addition to the add-on web service, each add-on has a web front-end, which is implemented with HTML and JavaScript. When the user starts an add-on, the host application loads the add-on UI into an \texttt{iframe} and displays it seamlessly as part of the user interface of the host application. There are two fundamentally different ways for the add-on UI to interact with the host application. It can either communicate locally with the host-application UI component or via the backend servers. In the latter case, the add-on UI usually connects to the add-on server in the cloud, which interacts with the host application server and accesses the user data through backend APIs that are not visible to the user.

\boldtitle{Access control.}
Cloud application vendors typically implement \emph{permission-based access control} for add-ons to limit their access to user data in the host application. Each add-on has a list of permissions which it requires to operate. The host application usually asks the user to explicitly approve the permissions when the user runs it for the first time or during its installation. This access control tends to be rather coarse grained, i.e.~the user has to grant all the requested permissions for either all user data or for a specific document. Furthermore, the add-on retains the permissions until the user uninstalls it.

\boldtitle{Marketplaces.}
Cloud application vendors often list their add-ons in an online \textit{marketplace}, from where the users can choose and install (i.e.~enable) any add-on for the applications. For instance, the G Suite marketplace~\cite{gsuitemarketplace} lists add-ons for Google applications such as Gmail and Google Docs. Usually, only a relatively small number of add-ons are provided by the application vendor itself, and the rest are built by third-party developers.


\section{XSS in vulnerable add-ons}
\label{sec:xss}

In this paper, we focus on \emph{non-malicious add-ons}. The add-ons are written by well-meaning developers who do not intend to cause harm but might not be security experts. Nonetheless, such add-ons can be vulnerable to external attacks, including XSS.

We have identified two types of XSS attacks against vulnerable add-ons (see Figure~\ref{fig:threat_model}):

\begin{enumerate}
  \item \textbf{Attack with shared workspace}: The attacker and the victim are colleagues, friends or remote collaborators, who use the same cloud application. The attacker shares a \textit{workspace} with the victim. The workspace concept varies depending on the host application, but it basically is any environment, such as a Google Docs document, through which changes made by one user are propagated to the others. The attacker injects malicious JavaScript code into the shared workspace. If the text would be visible to the user, it can be hidden with the usual techniques like using small font size or text color matching the background. When the victim enables a vulnerable add-on for the shared workspace and the add-on renders the attacker's input in an unsafe way, the injected script may become part of the web page in the add-on \texttt{iframe}, where it is executed by the victim's web browser. Thus, the attacker has performed an XSS attack on the victim.

  \item \textbf{Attack with outside input}: Some host applications accept external input such as messages from non-users. For example, if the host application is an email service (e.g.~Gmail or Outlook), the attacker can hide the malicious script in an email and send it to the victim. If the victim has enabled a vulnerable add-on to process emails, the injected script may again find its way to the add-on's \texttt{iframe} and be executed there like any JavaScript in that frame.
\end{enumerate}

The details of how the attacker injects the script into the shared workspace are naturally specific to the cloud application, to the add-on, and to the vulnerability that is being exploited. In any case, the root cause of the above attacks is that the vulnerable add-on routes the untrusted user input to JavaScript's data flow sinks in the add-on UI without sanitizing it. In particular, malicious data from the attacker can be executed if the add-on renders it as HTML rather than as text (with HTML element sinks e.g.~\texttt{document.write} or \texttt{document.body.innerHTML}). The attack would also succeed if the attacker's content is given as input to JavaScript methods such as \texttt{eval} and \texttt{setTimeout} which convert string input to code. (The latter type of mistakes are less likely though, because of developer awareness of the dangers of such functions.)

Compared to the traditional variants of XSS (see Section~\ref{sec:background_xss}), the attacks that we describe here are similar to \textit{stored XSS} because the attacker's malicious input is stored in the host application's data store. Also, they can be either server-side XSS or client-side XSS depending on whether the malicious data is processed by the add-on server or by the add-on UI.

\boldtitle{General consequences.}
With the ability to run arbitrary scripts in the context of the vulnerable add-on, the attacker can perform at least the following types of attacks:

\begin{itemize}
\item The attacker can access data on the add-on server through its APIs. In a microservice architecture, the add-on server is likely to have its own data storage.

\item The attacker may be able to access data in the host application with cross-domain messaging, or indirectly through the add-on server. The ability to do this depends on the design of the host application and its add-on APIs.

\item The attacker can spoof another user interface in the add-on \texttt{iframe} and trick the user into entering confidential data or credentials.

\item As in all XSS attacks, the attacker can access HTML5 APIs and request access to local resources, such as geolocation, or authorization to external resources owned by the victim user.
\end{itemize}

We will discuss the designs of different host applications and analyze what the attacker can gain in each case in Section~\ref{sec:architecture}. It is important to understand that the malicious script runs in the \texttt{iframe} with a different origin than the host application. Thus, it cannot access the DOM model of the host-application within the web browser or the cookies related to the host application. Instead, any access to host-application data has to be gained either through published APIs in the add-on server or with cross-origin access methods.


\section{Analysis of XSS consequences}
\label{sec:architecture}

To understand better what consequences there are to the execution of malicious code, we analyzed the add-on system architectures of three popular cloud application suites: Microsoft (MS) Office Online~\cite{officemarketplace}, G Suite~\cite{gsuitemarketplace}, and Shopify~\cite{shopifymarketplace}. This section presents our analysis in detail, and Table~\ref{tab:analysis_summary} at the end of the section summarizes the results.

\subsection{MS Office Online add-ons}
\label{sec:office}

\begin{figure}[tb]
  \centering
  \includegraphics[scale=0.6]{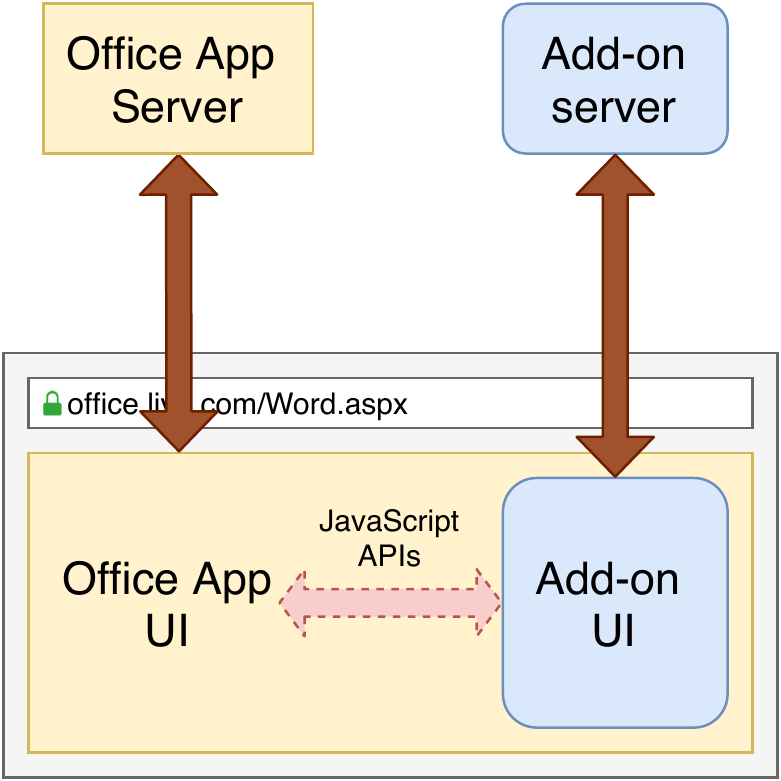}
  \caption{MS Office Online add-on architecture}
  \label{fig:office_architecture}
\end{figure}

MS Office Online is a cloud-based office suite, which includes popular office applications like Word, Excel, PowerPoint, and Outlook. The applications in the suite all have the same architecture, illustrated in Figure~\ref{fig:office_architecture}. The add-on UI is displayed inside the host application UI, which allows the user to interact with the add-on seamlessly when using the application. The add-on UI is contained in an \texttt{iframe} and has a different origin than the encapsulating application, which prevents it from directly accessing user data in the host application.

\begin{table*}[th]
  \centering
  \small
  \caption{MS Office Online add-on permission levels}
  \begin{tabular}{|l|l|p{11.8cm}|}
    \hline
    \textbf{Application} & \textbf{Permission} & \textbf{Description} \\
    \hline

    \multirow{4}{*}{Outlook}
     & Restricted & Read phone numbers, addresses, and URLs from the current item \\ \cline{2-3}
     & ReadItem & Read all properties of the current item \\ \cline{2-3}
     & ReadWriteItem & Full access to the current item \\ \cline{2-3}
     & ReadWriteMailbox & Full access to the mailbox\\
    \hline
    \hline

    \multirow{5}{*}{\parbox{2cm}{Word, Excel, \\ PowerPoint, OneNote}}
     & Restricted & Read/write settings of the add-in that are stored in the current document \\ \cline{2-3}
     & ReadDocument & Read only the text in the current document\\ \cline{2-3}
     & ReadAllDocument & Read everything in the current document, which includes text, formatting, links, graphics, etc. \\ \cline{2-3}
     & WriteDocument & Write to the user's selection in the current document. \\ \cline{2-3}
     & ReadWriteDocument & Full access to the current document \\ \cline{2-3}
    \hline

  \end{tabular}
  \label{tab:office_permissions}
\end{table*}

The add-on interacts with the host application on the client side via JavaScript APIs. Specifically, MS Office Online applications use \texttt{window.postMessage()}\cite{postmessage} for cross-origin messaging between the add-on's \texttt{iframe} and the parent application window.
The add-on can request different levels of access to the host-application data~\cite{officepermissions}, shown in Table~\ref{tab:office_permissions}. If the host application is Word, Excel, PowerPoint or OneNote, the add-on can only request access to the current document that the user is working on. Outlook add-ons, on the other hand, can request access not only to the current item (i.e.~email or compose form) but also to the user's whole mailbox. Outlook add-ons can also call \texttt{getCallbackTokenAsync()}, a special API that returns an \textit{access token} with the permission level of the add-on. The add-on UI running in the browser can send this token to the add-on server, which can use it to access the email server~\cite{outlookrestapis} on the add-on's behalf for a limited time.




\subsubsection{XSS exploits}
\label{sec:office_xss}

Let us consider how an attacker can exploit an MS Office Online add-on that is vulnerable to XSS. First, the victim needs to install the add-on. Then, depending on the add-on, the attacker can exploit the situation with either of the two attack vectors that we presented in Section~\ref{sec:xss}: it can inject malicious scripts into a document that is shared with the victim, or in the case of Outlook, it can send an email that contains the malicious scripts to the victim. Below, we consider what kind of access the attacker can gain to the user's data in the host application.

\boldtitle{Get the same level of access as the add-on.}
Because of the local messaging with the host application window, the attacker can access any resources that the add-on is permitted to access. Since the attacker's scripts run within the add-on's \texttt{iframe}, it is not possible for the host application to differentiate between malicious requests from the attackers and legitimate ones.

If the host application is Word, Excel, PowerPoint or OneNote, the attacker can access only the open document. This might not always be useful to the attacker because the document was originally shared with the attacker. However, the attacker can use the injected script as a backdoor to retain his access even after the victim revokes his legitimate access, e.g.~by making a personal copy of a form or template before filling it with confidential data. In Outlook, on the other hand, the attacker will gain full access to the victim's mailbox if the vulnerable add-on has the \emph{ReadWriteMailbox} permission. This means that the attacker can read all of the victim's emails and send emails on the victim's behalf.

\begin{figure}[tb]
  \centering
  \includegraphics[width=0.65\columnwidth]{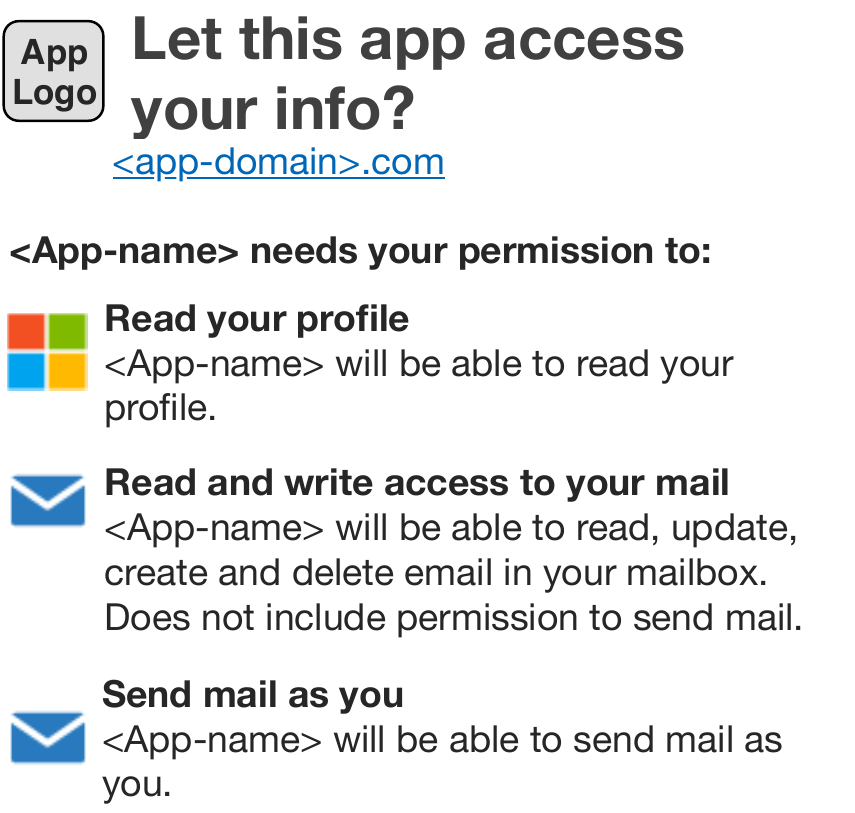}
  \caption{MS Office Online add-on authorization prompt}
  \label{fig:office_prompt}
\end{figure}

\boldtitle{Request an OAuth 2.0 token.}
As noted earlier, the ability to control the \texttt{iframe} enables the attacker to spoof parts of the application user interface, which makes it possible to trick the user in various ways, such as phishing for confidential data. We found, however, one specific trick that the attacker can play on MS Office Online users. Many add-ons act as the connectors between the MS Office Online applications and third-party web services built on the Azure platform. Such an add-on only provides a UI for the user to interact with the third-party backend server. Instead of acting like an add-on server, the server uses more powerful APIs for interacting with the host application. To obtain such access, the service provider must register an Azure application with the Microsoft identity platform~\cite{microsoftidentity}, and the user must authorize the application to access the necessary resources. The authorization is based on OAuth 2.0~\cite{rfc6749} as follows. The add-on displays a popup that shows information about the application including the name, logo and domain, as well as the permissions that the application is requesting, as illustrated in Figure~\ref{fig:office_prompt}. If the user agrees to authorize the application, the application will receive an \emph{access token}, which it can use to access the requested resources from anywhere.

Since the users of MS Office Online add-ons are already familiar with the OAuth authorization prompt, the attacker can exploit it to phish for access rights. First, the attacker creates an Azure application with the exact same name as the vulnerable add-on. This is possible because Azure applications do not need to have unique names. With the injected script, the attacker requests authorization for some of the user's resources. If the victim authorizes the attacker's application, the attacker can use the access token to access the victim's data in the host application from anywhere. The token is similar to the Outlook token discussed above but applicable to any of the MS Office Online applications. It is difficult for the victim user to judge whether a particular add-on should be granted an OAuth 2.0 token or limited to the add-on APIs.

\subsubsection{Case study: Translator for Outlook}
\label{sec:office_case}

We will use the "Translator for Outlook"~\cite{translatoroutlook}, an add-on developed by Microsoft itself, to demonstrate the exploits. As the name suggests, it is an add-on for the email service Outlook, which translates the user's emails to a selected language. The main workflow of the add-on is as follows.

\begin{enumerate}
  \item The user starts the add-on. The host application will display the add-on as a side panel in its UI.
  \item The user selects the language that she wants to translate the opening email to.
  \item The add-on translates the whole email to the selected language and displays the result.
\end{enumerate}

The problem with the add-on is that it renders the translated text as HTML without escaping the text first. As the result, if the attacker, which could be anyone on the Internet, sends an email that contains malicious scripts to the victim and the victim tries to translate it with the add-on, the scripts will be executed. The add-on, however, only has the \textit{ReadItem} permission. Thus, by exploiting the local messaging with the host application alone, the attacker will not be able to read the victim's mailbox or send emails on the victim's behalf. To gain such access rights, the attacker can use his malicious scripts to request an OAuth 2.0 token as we described in the previous section.


\subsection{G Suite add-ons}
\label{sec:gsuite}

G Suite is another office suite, which is developed by Google. Some well-known examples of the applications in the suite are Google Docs, Google Sheets, and Gmail.

Before going into details about how G Suite add-ons work, we need to understand the concept of Google APIs~\cite{googleapis}. They are a set of APIs that give programmatic access to many Google products, such as Google Maps and Google Drive. For a client (e.g.~a website) to access private user data using the APIs, it must be first attached to a Google Cloud Platform (GCP) project~\cite{gcpproject}. The client then needs to obtain an \emph{access token} with OAuth as follows. First, the user is redirected to the Google Authorization website, where the user must sign in with her Google account. The website then displays an authorization prompt showing the name of the GCP project and the permissions that the client is requesting. If the user grants the permissions, the Google Authorization server sends the access token to the client. Otherwise, the client receives an error.

\begin{figure}[tb]
  \centering
  \includegraphics[scale=0.6]{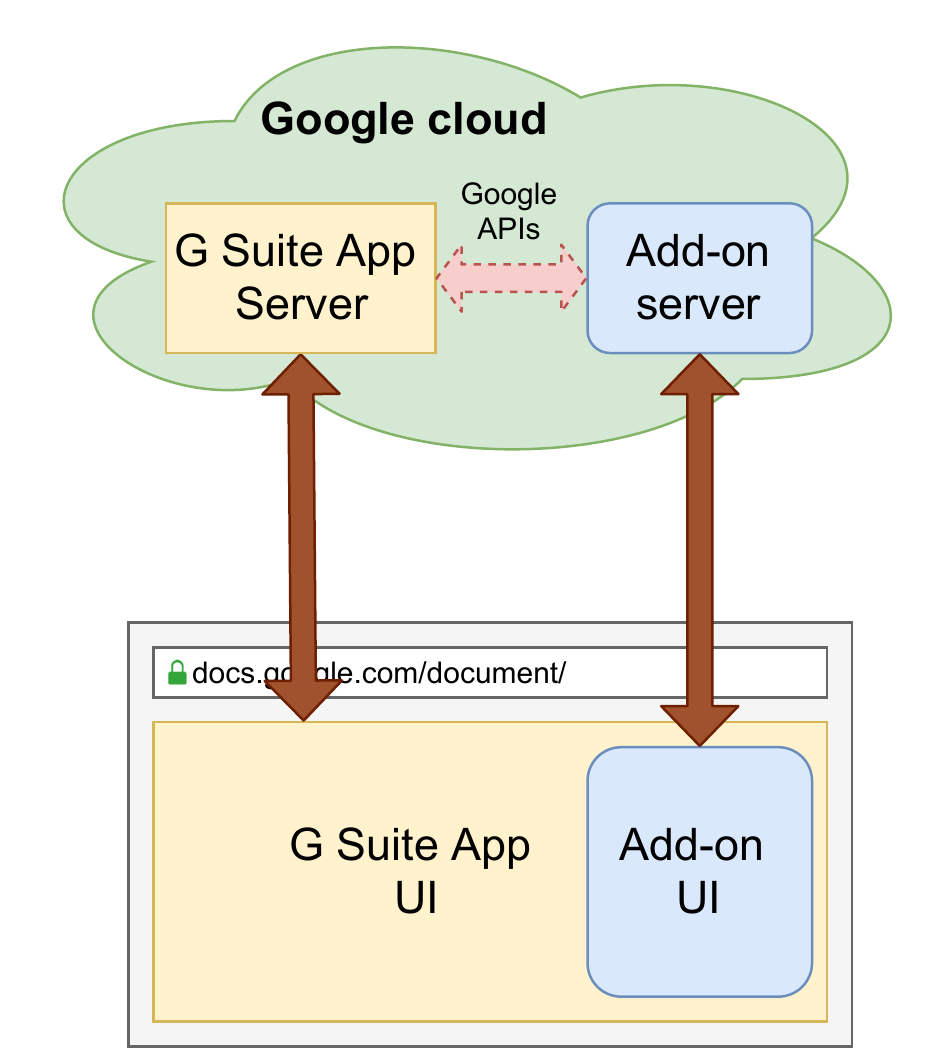}
  \caption{G Suite add-on architecture}
  \label{fig:gsuite_architecture}
\end{figure}

Figure~\ref{fig:gsuite_architecture} shows the architecture of G Suite add-ons. The main difference between them and MS Office Online's is that G Suite add-ons are completely hosted on the Google cloud. The add-on server is basically a Google APIs client that can directly interact with the user data. The add-on UI sends requests to interfaces defined by the add-on server, and the server implements the desired action on user data as well as returns responses. The server interfaces can only be accessed by add-on code that originates from the same server. One example is Translate~\cite{translateaddon}, a Google Docs add-on provided by Google. Its server has two main interfaces: one translates the user-selected text and returns the result, and the other replaces the text of the current selection with the translated text.

\begin{figure}[tb]
  \centering
  \includegraphics[width=0.75\columnwidth]{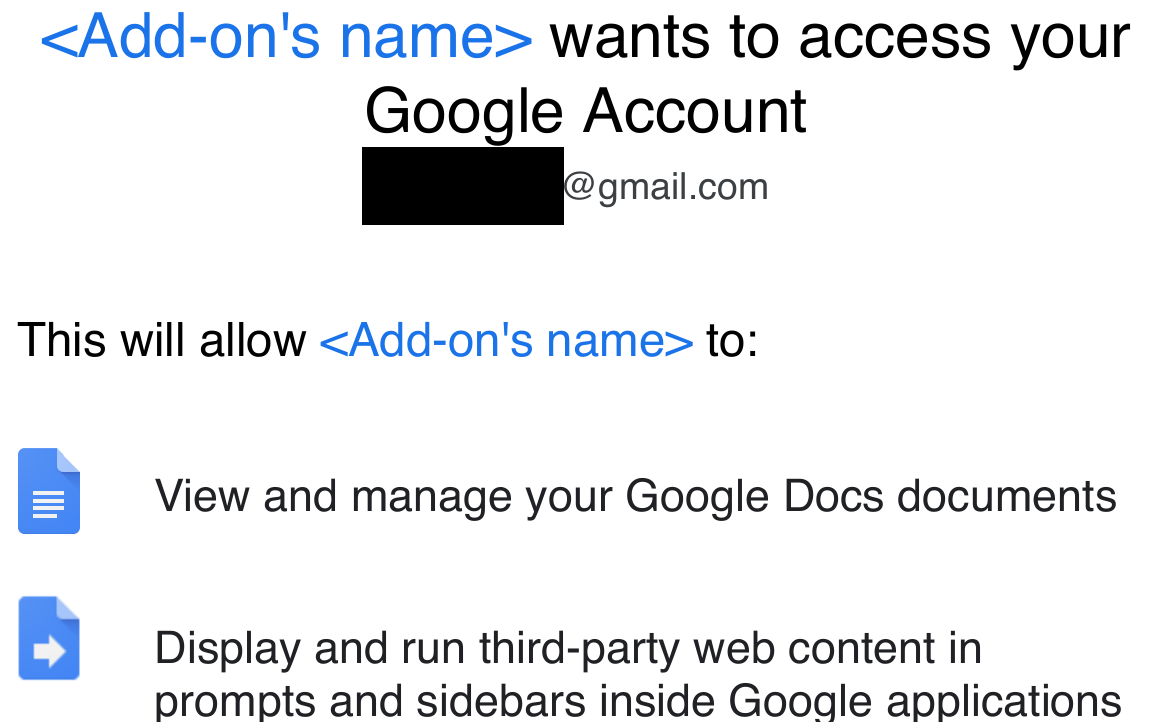}
  \vspace*{1em}
  \caption{G Suite add-on authorization prompt}
  \label{fig:permissions_prompt}
\end{figure}

Since the add-on server is a Google APIs client, it must be authorized before it can access the user's private data. This occurs when the user starts the add-on for the first time. Figure~\ref{fig:permissions_prompt} shows a typical authorization prompt. If the user approves, the add-on server obtains an access token with the requested permissions.

G Suite add-ons can request permissions to access user data in any G Suite applications. For example, a Google Docs add-on can request permissions to send emails from the user's Gmail account. While this cross-application permissions model makes the add-ons flexible and powerful, malicious add-ons could exploit it to gain access to a wide range of user data by, for example, requesting permissions to ``Read, compose, send, and permanently delete all your email from Gmail'' and ``See, edit, create, and delete all of your Google Drive files''. To mitigate such threats, Google performs manual verification of add-ons that request sensitive permissions~\cite{oauthclientverification} to ensure that the add-ons comply with the Google API User Data Policy~\cite{googledatapolicy}.


\subsubsection{XSS exploits}
\label{sec:gsuite_xss}

Next, we consider what kind of access a successful XSS attacker can get to user data.

\boldtitle{Get the same access as the add-on.}
At first glance, since the host application window does not accept local messages from the add-on UI, it appears that the XSS attacker cannot access the victim's data. However, there is a very common Google API used by the add-on UIs that allows the attacker to bypass this limitation: Google provides the Picker API~\cite{googlepicker} for the user to select a file or folder that is stored in Google servers. Like any other Google APIs, the Picker API requires an access token to operate. The add-on servers commonly create an interface by which the add-on code running in the browser can obtain a copy of the server's token; this is even recommended practice~\cite{googlepickerexample}. The server's token, however, is not limited to the Picker API. Now, the injected XSS code can also request the access token, and thus it will gain the same permissions to the user's data as the add-on server has.

\boldtitle{Request an OAuth 2.0 token.}
If the vulnerable add-on does not use the Picker API, the attacker can turn the injected script into a Google APIs JavaScript client and request an OAuth 2.0 token from the user. In that case, an authorization prompt is displayed to the victim in the same way as when an add-on requests permissions in its first run (see Figure~\ref{fig:permissions_prompt}). While the attacker must use his own GCP project for his malicious client, it could choose a name for the project that matches the add-on's name.

This is similar to the phishing exploit that we presented for Office Online add-ons in Section~\ref{sec:office_xss}. We believe that this attack will have a high success rate because G Suite users are already familiar with the authorization prompt, and the victim might think that the add-on has been updated and needs new permissions.


\subsubsection{Case study: Form Ranger}
\label{sec:gsuite_case}

Form Ranger~\cite{formranger} is an add-on for Google Form with the most number of users. Google Forms is an online service that helps collect information from users via surveys and quizzes, and the add-on allows automatically populating a form with data from any spreadsheet in the user's Google Drive. The main workflow of the add-on is as follows.

\begin{enumerate}
  \item The user starts the add-on. The host application will display the add-on as a side panel in its UI.
  \item The add-on shows the list of questions in the form. For each question, the add-on allows the user to populate the answer options with data from a spreadsheet.
  \item \label{step:formranger_picker} The user selects a question, and the add-on displays a list of all spreadsheets that are stored in the user's Google Drive. Note that the Picker API is used here.
  \item \label{step:formranger_vuln} The user selects a document in the list, and the add-on allows selecting which sheet and which column in the sheet that the user wants to import data from. A preview of the data in the column is also displayed.
  \item The user select a sheet and a column, and the add-on populates the question with the data in the column.
\end{enumerate}

At this point, the readers can probably guess in which step the add-on is vulnerable: in Step~\ref{step:formranger_vuln}, the add-on does not filter or escape the data from the selected spreadsheet document and renders it as HTML. Thus, if the attacker has access to the document (e.g. the attacker and the victim are collaborators), it can hide malicious JavaScript code in the part of the document that will be used as the form inputs and the code will be executed when the victim uses the add-on. Since the add-on uses Picker API (Step~\ref{step:formranger_picker}), the attacker can steal its access token and gain all of its permissions, which include the following: ``See, edit, create, and delete all of your Google Drive files'', ``See, edit, create, and delete your spreadsheets in Google Drive'', ``View and manage your forms in Google Drive'', and ``Send email as you''.

We can see that the add-on has more permissions that it actually needs. For example, it does not need the ability to send emails on the user's behalf. It also does not need access to all of the Google Drive files; read access to the spreadsheets only should be sufficient. Because of these unnecessary permissions, the attacker will be able to access all of the victim's Google Drive files and send emails on the victim's behalf. If the attacker wants to gain even more access rights (e.g. read the victim's emails), it can phish to obtain an OAuth 2.0 token as we described in the previous section.



\begin{table*}[th]
  \centering
  \caption{Analysis summary}
  \def\arraystretch{1.2}
  \begin{tabular}{|p{1.6cm}|p{3.2cm}|p{2.2cm}|p{2.5cm}|p{6.1cm}|}
    \hline
    \textbf{Application}
    & \textbf{Add-on's data access}
    & \textbf{Add-on's logic}
    & \textbf{Attack vectors}
    & \textbf{Exploits} \\ \hline

    MS Office \newline Online
        & Host application only
        & In frontend
        & Shared workspace \newline Outside input
        & - Get the same level of access as the add-on \newline
          - Request an OAuth 2.0 token
        \\ \hline

    G Suite
        & Any applications
        & In backend
        & Shared workspace \newline Outside input
        & - Get the same level of access as the add-on only if the Picker API is used \newline
          - Request an OAuth 2.0 token
        \\ \hline

    Shopify
        & Host application only
        & In backend
        & Shared workspace
        & - Install a malicious add-on
        \\ \hline

  \end{tabular}
  \label{tab:analysis_summary}
\end{table*}


\subsection{Shopify add-ons}
\label{sec:shopify}

Shopify is an e-commerce platform with which small merchants can create online shops. It offers services such as payment, marketing, and customer engagement. Each shop is managed through a \emph{web admin interface}, on which the owner can access the built-in services of the platform to, for example, add new products to the shop, engage with users, or manage orders. Shopify add-ons integrate third-party services into this admin interface.

\begin{figure}[tb]
  \centering
  \includegraphics[scale=0.6]{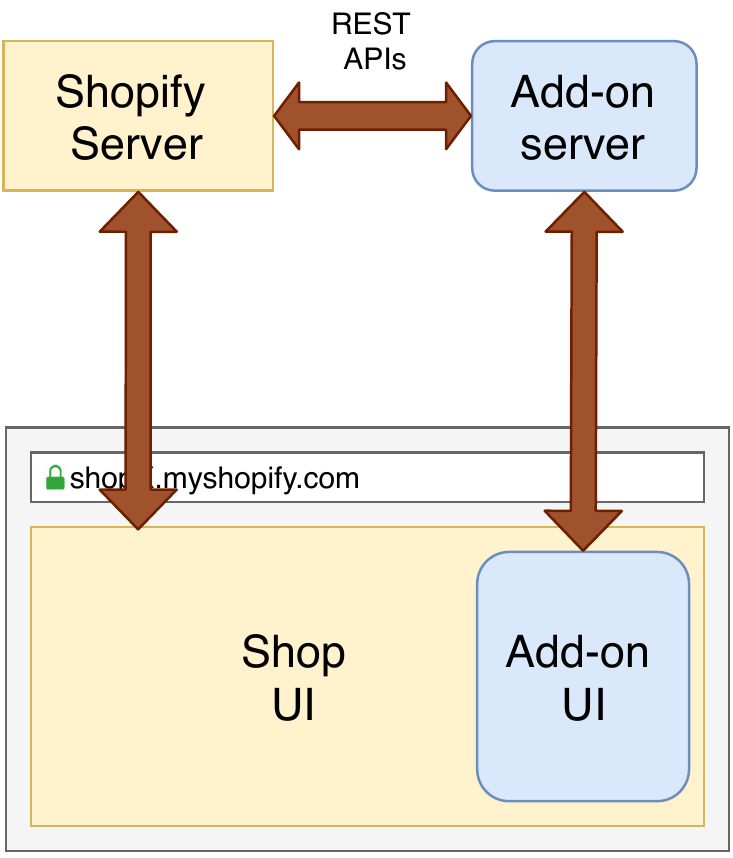}
  \caption{Shopify add-on architecture}
  \label{fig:shopify_architecture}
\end{figure}

The add-on architecture is illustrated in Figure~\ref{fig:shopify_architecture}. The add-on server runs in the cloud and accesses the shop data with the Shopify REST APIs over HTTPS. Typically, the add-on server is authorized to access the shop server with an OAuth 2.0 \emph{access token}, which it obtains as follows. When the user starts the add-on for the first time, the shop's admin interface shows an authorization prompt with a list of the data that the add-on requests to access (e.g.~orders, products). If the user agrees, the Shopify server sends an access token to a pre-registred endpoint on the add-on server. The add-on server can then use the token to access the requested data. Later, the add-on can renew the access tokens without prompting the user.

The add-on UI is embedded into a menu in the shop's admin interface. Shopify provides a set of local JavaScript APIs~\cite{shopifyappbridge} for the add-on UI to perform resource-picking operations. Specifically, the add-on UI communicates with the shop's admin interface via \texttt{window.postMessage()}, and the admin interface asks the user to pick resources of the requested type. The admin interface then returns the result data objects to the add-on.

A shop on Shopify can have an owner and multiple staff members. The shop owner has full access to the shop and can grant the other staff members access to some (or all) admin features including add-ons~\cite{shopifypermissions}. The view through an add-on's UI is based on the intersection of the member's and the add-on's permissions.

\subsubsection{XSS exploits}
\label{sec:shopify_xss}

With Shopify, the attacker must be a staff member who is able to inject malicious scripts into the shop resources (i.e.~the ``shared workspace'' attack vector). We found that Shopify prevents including HTML or scripts in customer information, but it is possible to inject scripts into the product and order descriptions. These scripts will be executed when other staff members use a vulnerable add-on that renders them in an unsafe way.

\boldtitle{Install a malicious add-on. }
The Shopify add-on architecture has the good points of both MS Office Online and G Suite. In particular, the access token is sent directly to the add-on server and, thus, is not accessible to the add-on UI by default. Also, the local JavaScript APIs allow the add-on UI to perform resource-picking operations without exposing the shop data because it is processed in the admin interface.
Nevertheless, it is possible for the attacker to trick the shop owner or another authorized victim into installing a malicious add-on. Unlike MS Office Online and G Suite applications, whose add-ons can only be installed from the respective marketplaces, Shopify users can initiate the installation process of a new add-on just by visiting a URL. Thus, the attacker can create a malicious add-on with a name similar to that of the vulnerable add-on and initiate its installation from the injected script. The victim might think that the add-on has been updated and it needs to be authorized again. Depending on the extent of the permissions granted, the malicious add-on can then access any shop data.

\subsubsection{Case study: Order Printer Pro}
\label{sec:shopify_case}

Order Printer Pro~\cite{orderprinterpro} is one of the most popular add-ons in the ``Orders and shipping'' category on the Shopify marketplace. The add-on allows its users to quickly create order-related documents (e.g. invoices, return forms) as well as printing or delivering them to the customers. The workflow for printing an order is as follows.

\begin{enumerate}
  \item The user starts the add-on from the admin interface. The host application will display the add-on as a part of its UI.
  \item The user selects the ``Orders'' menu of the add-on UI. The add-on will show a list of the orders of the shop.
  \item The user selects the order that she wants to print. The add-on will display the order and action buttons that allow the user to print the order or export the order as a PDF document. The displayed information includes the customer's shipping address, the items in the order, and the notes that are written by the staff members if any.
  \item The user clicks on the ``Print'' button and the order will be printed.
\end{enumerate}

The problem with the add-on is that when it displays the order's information, it does not render the notes in a safe way. Consequently, if the notes contain JavaScript snipets, the snipets will be executed. As we mentioned in the previous section, Shopify does not prevent including HTML or scripts in the field. Thus, any staff member that have write access to the orders of the shop can perform XSS on other staff members, including the shop owner.

\section{Empirical analysis}
\label{sec:EmpAnalysis}

To find out how widespread the XSS issue in cloud-application add-ons is in the wild, we conducted an empirical analysis by looking for vulnerable add-ons in the marketplaces of the three selected cloud application suites. In this section, we first present our approach for shortlisting the add-ons, followed by the methodology that we used for systematically finding vulnerabilities in the shortlisted add-ons. Finally, we show the analysis results.

\subsection{Shortlisting add-ons for analysis}
\label{sec:shortlisting}

We consider only free add-ons from the marketplaces. Table~\ref{tab:marketplaces_stats} shows the number of such add-ons as of August 2019. Note that Microsoft and Shopify officially use the terms \textit{add-in} and \textit{app}, respectively.

\begin{table}[h]
  \centering
  \small
  \caption{Add-on marketplaces (August 2019)}
  \begin{tabular}{|l|l|c|}
    \hline
    \textbf{Marketplace} & \textbf{Terminology} & \textbf{Available free}\\
    \hline

    MS Office Online & add-in & 1150\\ \hline
    G Suite & add-on & 1180 \\ \hline
    Shopify & app & 1265 \\ \hline

  \end{tabular}
  \label{tab:marketplaces_stats}
\end{table}


\newcommand{\fixing}{Fixing}
\newcommand{\fixed}{Fixed}
\newcommand{\noresp}{No response}

\begin{table*}[tbh]
  \centering
  \caption{Vulnerable add-ons found in our empirical analysis}
  \begin{tabular}{|l|p{1.5cm}|p{6cm}|c|c|}
    \hline
    \textbf{Marketplace} & \textbf{Selection criterion} & \textbf{Vulnerable add-ons} & \textbf{Attack vector} & \textbf{Status} \\
    \hline

    MS Office Online  & Popular & Translator for Outlook& Outside input & \fixing \\ \cline{3-5}
                      &         & GIGRAPH - Network Visualization& Shared workspace & \noresp \\ \cline{2-5}

                     & Random  & Duplicate Remover& Shared workspace & \fixing \\ \cline{3-5}
                     &         & Bubbles& Shared workspace & \noresp \\ \cline{3-5}
                     &         & Radial Bar Chart& Shared workspace & \noresp \\ \cline{3-5}
                     &         & Excel to JSON& Shared workspace & \noresp \\ \cline{3-5}
                     &         & WritingAssistant& Shared workspace & \noresp \\ \cline{3-5}
                     &         & Excel to SMS& Shared workspace & \noresp \\ \cline{3-5}
    \hline \hline

    G Suite & Popular & Form Ranger& Shared workspace & \fixing \\ \cline{3-5}
            &         & Flubaroo& Shared workspace & \noresp \\ \cline{3-5}
            &         & autoCrat& Shared workspace & \fixing \\ \cline{3-5}
            &         & formMule - Email Merge Utility& Shared workspace & \fixing \\ \cline{3-5}
            &         & docAppender& Shared workspace & \fixing \\ \cline{3-5}
            &         & Grackle Sheets& Shared workspace & \noresp \\ \cline{3-5}
            &         & Sheetgo& Shared workspace & \fixing \\ \cline{2-5}

            & Random  & Form Duplicates& Shared workspace & \noresp \\ \cline{3-5}
            &         & Bulk Sheet Manager& Shared workspace & \noresp \\ \cline{3-5}
            &         & rosterSync - Teacher Edition& Shared workspace & \fixing \\ \cline{3-5}
            &         & Notifications for Forms& Shared workspace & \noresp \\ \cline{3-5}
            &         & Text gBlaster (SMS Texting)& Shared workspace & \noresp \\ \cline{3-5}
            &         & Mail Merge& Shared workspace & \noresp \\ \cline{3-5}
            &         & Response Editor& Shared workspace & \noresp \\ \cline{3-5}
            &         & Doc Variables & Shared workspace & \fixed \\ \cline{3-5}
    \hline \hline

    Shopify & Popular & Order Printer Pro& Shared workspace & \noresp \\ \cline{2-5}
            & Random  & ShipHero Fulfillment& Shared workspace & \fixing \\ \cline{3-5}
            &         & Simple Admin& Shared workspace & \noresp \\ \cline{3-5}
            &         & ShipRelay Fulfillment& Shared workspace & \noresp \\ \cline{3-5}
            &         & Ship Systems 3D Box Packing& Shared workspace & \noresp \\ \cline{3-5}
    \hline

  \end{tabular}
  \label{tab:vulnerable_addons}
\end{table*}

We selected 100 free add-ons from each of the three marketplaces --- 50 popular ones that are likely to have many users, and another 50 selected randomly--- resulting in a total of 300 add-ons for the study. The criteria used for shortlisting the add-ons are as follows.

\boldtitle{MS Office Online}: We focused on add-ons for the following applications in the suite: Word, Excel, OneNote, Outlook, and PowerPoint. The remaining applications either have no add-ons or the add-ons are available only to domain users. Since the marketplace does not show the number of users for each add-on, we shortlisted the top 50 add-ons based on the number of reviews.

\boldtitle{G Suite}: The applications available to individual users include Gmail, Docs, Sheets, Slides, Forms, Drive, and Calendar. We excluded Drive or Calendar because their add-ons only add new menus to the application UI, which means that they do not have any client-side code, leaving no room for the XSS attacker. From the remaining applications, we shortlisted 50 add-ons with the highest number of users.

\boldtitle{Shopify}: We first sorted add-ons in the marketplace by the ``Most installed'' option, and then we shortlisted the first 50 add-ons from the result. Note that the marketplace does not show the number of installations of each add-on; instead, it shows only the rating and the number of reviews.

\subsection{Analysis methodology}
\label{sec:methodology}

We manually analyze the selected add-ons with \emph{black-box testing}. There are several challenges with automatically verifying whether an add-on would cause XSS vulnerability.
First, state-of-the-art automated techniques for detecting XSS vulnerabilities typically work only with client-side XSS~\cite{steffens2019don,lekies201325,melicher2018riding,saxena2010flax}, while the nature of the XSS attacks against cloud-application add-ons vary depending on both the host application and the implementation of the add-on. Specifically, the vulnerabilities of MS Office Online add-ons are always on the client side because the attacker's malicious input is propagated to the add-on via local messaging (i.e.~\texttt{window.postMessage()}) and processed by the add-on UI. The vulnerabilities of G Suite and Shopify add-ons, on the other hand, can be either on the client side or on the server side depending on whether the add-on server returns the attacker's raw input to the client or it renders the data beforehand.
Second, unlike traditional web applications where a significant portion of the workflows can be analyzed by loading every available URL in the websites with a browser, most functionality of an add-on can only be triggered by user actions. In addition, there is no single standard method to invoke an add-on. Some start by the user clicking and selecting the add-on from a host-application menu, while the others require further user interaction to reach a point where the XSS code may be executed.
Third, most automated XSS detection approaches in the literature do not involve user logging in. However, many add-ons act as the connectors between their host applications and third-party web services; thus, they require their users to create an account and log in before they can be used so that they keep track of the user data. The registration processes vary greatly, making them difficult to be done without manual user interaction.
Because of these reasons, we leave automatic detection of XSS vulnerabilities in cloud-application add-ons to future work.

\boldtitle{Black-box testing.}
For each add-on, we first installed it and tried to understand its features. We then created a test item for each of the target applications as follows.

\begin{itemize}
  \item For document editing applications in MS Office Online and G Suite (e.g.~Word, Excel, Google Docs, Google Sheets), the test item was a corresponding document. To make sure that every source of data that the add-on was going to process contained JavaScript code, we added simple JavaScript snipets to every part of the document. For instance, if it was a spreadsheet, we included JavaScript code in the name of each sheet, the heading of each column, and some cells in each column. An example of the snippets is \texttt{<script>console.log(``Pwned!!!'')</script>}.

  \item For email applications (i.e.~Outlook, Gmail), the test item was an email that we sent to ourselves. Like with document editing applications, we inserted scripts to the subject and every part of the email's body.

  \item For Shopify: We first created a test shop. As we mentioned in Section~\ref{sec:shopify_xss}, it is possible to injects scripts only into the products and orders. Thus, we created several test products and orders in the shop, and we added scripts into every possible places, such as the name and the description of the items.
\end{itemize}

After creating the test item, we tried all functions of each add-on on the test item and looked for workflows that involve rendering the content of the item. We also added more custom JavaScript code in a number of cases because some add-ons only considered data in specific formats. For example, with Doc Variables~\cite{docvariables} --- an add-on that allows defining and using variables in Google Docs documents --- we had to insert our script into a variable's definition, which was in the format: \texttt{\$\{variable\_name\}}. If any of the injected script snippets were executed, we concluded that the add-on was vulnerable to XSS.

While our analysis was manual, we believe that it is sufficient to find most, if not all, XSS vulnerabilities in the selected add-ons. The main reason is that, compared to standalone web services, add-ons are usually quite simple with a relatively small number of features. The number of places where the user can inject scripts is also limited. Thus, it is not difficult to understand and manually test all the workflows of an add-on with the black-box testing method. However, while our methodology is sufficient to produce results that, in our opinion, need wider attention, it is clearly not practical for those who want to perform large scale analyses of cloud-application add-ons. A more complete and efficient approach is needed in the future for such purpose.

\subsection{Results and responsible disclosure}
\label{sec:results}

We found 28 vulnerable add-ons among the 300 analyzed ones, which is around 9\%. The result indicates that XSS vulnerabilities are common in cloud-application add-ons today. Table~\ref{tab:vulnerable_addons} shows the names of the add-ons, the attack vectors, and whether the add-ons have been fixed. Among the three marketplaces, only the G Suite shows the number of users of each add-on, and the vulnerable add-on with the greatest number of users in the marketplace, Form Ranger, had roughly 7,8 million users as of August 2019. The most popular vulnerable add-ons in the Microsoft Office Online and Shopify marketplaces are Translator for Outlook and Order Printer Pro, respectively. The former had 1772 reviews, and the latter had 371 reviews.

We can see that the vulnerability rate in the set of popular add-ons appears lower in all the three marketplaces. This could be because popular add-ons are more likely to be written by more experienced developers. Also, it seems that add-ons that are vulnerable to outside input are rare. Specifically, only one add-on in our study is vulnerable. We hypothesize that add-on developers are more familiar with threats from outside input (i.e. emails in this case) than those from shared workspace.

We have disclosed the vulnerabilities to all of the add-on developers that we were able to contact. We also provided guidance on how the security bugs could be fixed (see the solutions for add-on developers in Section~\ref{sec:defenses}). At the moment of writing, of the 28 add-on teams/developers that we have contacted, 1 has acknowledged and fixed their add-on, 9 has acknowledged the vulnerabilities but are still working on the fixes, and the others have not responded to us. We also discussed the problem of the Picker API with Google. They confirmed the problem and said that they would take it into account in the next version of their add-on system.

\section{Defenses}
\label{sec:defenses}

In this section, we discuss what the add-on developers and cloud application vendors can do to defend against the XSS attacks caused by vulnerable add-ons.

\subsection{Solutions for add-on developers}
\label{sec:defenses_developers}

To prevent XSS, add-on developers should not add untrusted data to the add-on UI as HTML because it can contain malicious JavaScript code. We discuss some practices that they can follow below.

\boldtitle{Coding practices.}
Secure coding practices to prevent XSS are a common topic in web security literature~\cite{microsoftaddinsecurity,owaspxssprevention,grossman2007xss}. A straightforward way to prevent XSS in cloud-application add-ons is to always render user input as text instead of HTML. Instead of the \texttt{innerHTML} property, the developer should use the \texttt{innerText} and \texttt{textContent} properties to insert text. With jQuery, the \texttt{.text()} method should be used instead of the \texttt{.html()} method.

In general, user data should not be interpreted as web application code. However, in the rare cases where it is necessary to render untrusted HTML as part of the add-on UI, the developer needs to properly validate and escape on the input first. On the server side, most web frameworks have built-in functions for such tasks. On the client side, JavaScript methods like \texttt{.toStaticHTML()} can be used to remove dynamic HTML elements and attributes in the user data before rendering it. We refer to~\cite{owaspxssprevention} for a detailed guidance on how to escape characters to prevent XSS.

\boldtitle{Security enforcement.}
Since add-ons are basically web services, add-on developers could implement a Content Security Policy (CSP)~\cite{stamm2010reining} to enforce some defenses on their add-ons. An extreme policy is to ban execution of all inline scripts (e.g.~\texttt{<script>} tags, inline event handlers). With such policy, even if the attacker managed to insert malicious scripts to the add-on UI, the scripts would not be executed. Only JavaScript code that are in separate \texttt{.js} files and loaded from trusted servers that the developers have whitelisted are allowed to run.

However, completely prohibiting inline scripts is not always ideal because legitimate inline scripts are preferred for various tasks. For example, event handlers are usually implemented in inline scripts. Moving inline event handlers to separate \texttt{.js} files cannot be done by simply copying and pasting the code; instead, they need to be rewritten with DOM APIs. Fortunately, CSP can be used in a less extreme way to avoid the hassle. Specifically, hash-based or nonce-based policies can be implemented so that inline scripts with pre-registered hash or nonce values are allowed to execute.

Add-on developers should also minimize the permissions that their add-ons request. As we can see from the case of the Form Ranger add-on (Section~\ref{sec:gsuite_case}), the unnecessary permissions that the add-on has enable the XSS attacker to steal all of the victim's Google Drive files as well as sending emails on the victim's behalf.

There are also other generic practices that help to defend against XSS~\cite{owaspxssprevention}. For example, the \texttt{HTTPOnly} flag of session cookie and any custom cookies that are not accessed by any JavaScript code should always be set. Also, many web frameworks provide automatic escaping functionality~\cite{angularjsescaping,gotemplates}, which should be used whenever possible.

\boldtitle{XSS detection.}
Add-on developers can also utilize the method that we used for our empirical analysis (Section~\ref{sec:methodology}) to check whether their add-ons are vulnerable. Specifically, they can create similar test items as we did and write unit tests to continuosly check for XSS vulnerabilities during their development cycle. While our method does not scale well, it should be sufficient for testing individual add-ons.


\subsection{Lessons for cloud-application vendors}
\label{sec:defenses_vendors}

In Section~\ref{sec:architecture}, we analyzed the designs of three popular add-on systems and what an XSS attack can achieve in them. This section presents the lessons that we learned from the analysis about design choices and their impact on the security of add-ons. We hope that the lessons would be useful for any cloud-application vendors which are developing or improving their add-on systems.

\boldtitle{Harden the add-on \texttt{iframe}.}
The add-on UI is contained in an \texttt{iframe}, and by default, the code in the \texttt{iframe} can call browser APIs to request access to features on the local device, such as geolocation, microphone, and camera. The host application should restrict the browser features which the add-on \texttt{iframe} can access, which can be done by setting the \texttt{allow} and \texttt{sandbox} properties of the \texttt{iframe}~\cite{mozillaiframe}.

At the moment, there is an experimental feature in the Chrome and Opera browsers that allows the host application to enforce a CSP policy on add-ons. Specifically, the browsers add a new property to the \texttt{iframe} elements, namely \texttt{csp}, which can be used to specify a policy which the embedded page must enforce upon itself~\cite{w3ccspiframe}. If this property becomes a standard feature of browsers, it gives cloud-application vendors control over the CSP policy in their add-ons. However, we observe that most of the add-ons in our analysis did not use CSP. Thus, deploying such restriction would not be an easy task for the vendors because a meaningful CSP, such as allowing only inline scripts whose hash or nonce match a specified value to run, would break many add-ons in their marketplaces.

\boldtitle{Implement add-on logic in the add-on server, not in client-side JavaScript.}
Cross-origin messaging within the browser, as used in MS Office Online, enables low-latency access from the add-on to user data that is available in the main application UI. However, this access is only controlled by relatively coarse-grained generic permissions, such as those in Table~\ref{tab:office_permissions}. We can see that such access control is not useful when defending against XSS attackers because their malicious scripts are executed in the context of the add-on. Therefore, the add-on logic should be implemented on the server side, as G Suite and Shopify have done.
The add-on server would act a a layer of isolation between the client-side script and user data in the cloud application in two ways. First, the add-on server defines a limited, purpose-specific interface through which all access to user data has to go. Second, the add-on server implements business logic that further filters the kinds of read and write operations that are passed on to the cloud application. Both mechanisms act as filters between the potential XSS code that has taken over the add-on \texttt{iframe} in the browser and the user's data.

\boldtitle{Filter scripts in user input.}
The host application vendors should think thoroughly about the types of user input that their applications need to receive in a shared workspace and filter for unwanted types. For example, Shopify deals with shop resources such as products and orders, which are relatively structured data. Thus, the developers know where HTML or scripts should not appear. The G Suite and MS Office Online applications are more problematic in this respect because the input to them is mostly documents, where legitimate HTML and scripts can appear in unexpected places.

\boldtitle{Do not share access tokens to delegate all your permissions.} OAuth 2.0 tokens are bearer tokens, and anyone in possession of the token can use it for resource access. This can lead to unsafe coding practices where too powerful tokens are delegated to unsafe places --- as we saw in the case of the Picker API in G Suite. Instead of sharing its access token with the client side, the add-on server should mediate the access. Where tokens need to be shared, they should only convey the absolute minimum permissions needed by the client side. In the case of the Picker API, for example, the add-on server should delegate to the UI a restricted token that can only be used to list specific files in the user's Google Drive.

\boldtitle{Avoid asking user consent at runtime.}
Relying on user judgment when authorizing add-ons to access user data may not be as good an idea as it first seems. In particular, prompting the user for consent at runtime conditions the user to answering yes to every such prompt, including ones from injected malicious code. Thus, it may be better to ask for user consent only in a separate UI where the user installs or upgrades add-ons. On the negative side, that prevents document-level access control.

\section{Discussion and future work}
\label{sec:discussion}

Since cross-site scripting is a well-known vulnerability, prudent engineering practices have been developed to prevent such mistakes~\cite{owaspxsscheatsheet}. On one hand, developers are aware of the need to filter untrusted input, and on the other, cloud application vendors have developed platforms and toolchains that make their products immune to most types of code injection. Nevertheless, the problem has not been completely solved. On one hand, attackers find new ways of bypassing the defenses and, on the other, the speed of software development makes it difficult for threat analysis and defenses to stay up to date. This paper is doing its part to catch up with the development in one key area of modern software.

We have confirmed by experiments that the vulnerabilities described in this paper are real and exploitable. There are, however, some additional practical considerations that a real-world attacker would face. The attacker needs to know which vulnerable add-on the victim is using, and the victim has to enable the add-on on the shared document. Thus, a successful attack probably requires a vulnerable add-on that users regularly invoke on large classes of documents which they are reading or editing. Translator and writing-assistant add-ons could meet these criteria. Add-ons that fix problems in data, such as duplicate removers, could even be installed by the victim when they receive a document with the matching problem.



In addition to the XSS vulnerabilities in the add-ons, another serious issue discovered in this paper is the way OAuth 2.0 tokens are used in the Picker API, and the powerful exploits that it enables for the XSS attacker. The Picker API documentation has educated developers to use a design pattern where an add-on server shares its OAuth 2.0 with a client-side script. Even if the Picker API itself is replaced with a safer solution, this unsafe software pattern might persist among developers.

For these reasons, some further measures may be needed to prevent unsafe use of the access tokens. G Suite has the advantage that the add-on server runs in the Google cloud, and the access token is handled by the third-party code only in special cases, which could be monitored and blocked. Also, host applications could relatively easily reject tokens that come from somewhere other than the authorized add-on server. Such restrictions on the token usage would, however, take away the convenience and flexibility of bearer tokens that has made them popular with developers. Indeed, the designers of OAuth 2.0 have intentionally not built in support for channel binding, such as binding the token to a specific client address. In conclusion, there are ways of mitigating the threat caused by unsafe delegation of access tokens to client-side scripts, but it might take some time for the problem to go away entirely.

Overall, we  hope that this paper will attract more attention to the area of cloud-application add-ons since further work is clearly needed. There might be other attack vectors that allow the attacker to exploit non-malicious add-ons. Analyzing threats from malicious add-ons could also be an interesting area for future work.


\section{Related work}
\label{sec:related}

In this section, we survey related literature about XSS attacks and security analysis of add-on ecosystems outside the domain of web applications.

\boldtitle{XSS vulnerabilities. } XSS has been one of the most common and harmful vulnerabilities in web applications. In spite of the availability of detection and defense mechanisms and changes in the architecture of web applications, XSS remains a prevalent problem~\cite{steffens2019don,chen2019comparison}. Security research literature on XSS includes a comprehensive overview~\cite{gupta2017cross}, detection mechanisms~\cite{wassermann2008static, johns2008xssds, ismail2004proposal}, as well as preventive and defense solutions~\cite{ter2009blueprint,vogt2007cross,bisht2008xss,nadji2009document,van2009noncespaces,kirda2006noxes}.

The rich literature on defenses again XSS includes both client and server-side solutions. Many of them can help to defend against the attacks presented in this paper.

Client-side solutions involve sanitizing user input before it is sent to the server. However, distinguishing between trusted and untrusted content and filtering out any malicious scripts are challenging tasks. This is why the sanitation of web pages is sometimes outsourced to the browsers~\cite{ter2009blueprint} or to web firewalls that run on the client PC~\cite{kirda2006noxes,ismail2004proposal}. Even though the XSS attacks occur on the client-side, solutions often involve server side mechanisms. For example, in the solution of Gundy et al.~\cite{van2009noncespaces}, the potentially vulnerable website delivers a XHTML document with randomized namespace prefixes and a policy to the client, and the client accepts only documents that comply with the policy.

Taint checking is a popular server-side protection mechanism, where the input originating from untrusted sources is flagged as potentially malicious and subjected to further scrutiny (e.g.~sanitizing the input). The same techniques can also be employed on the client side if combined with static analysis of the input strings~\cite{vogt2007cross, wassermann2008static}. There are server side solutions, for example passive monitoring of the  HTTP traffic~\cite{johns2008xssds} or by dynamically comparing HTTP responses with a pre-defined response~\cite{bisht2008xss}.

\boldtitle{Add-ons outside web applications.}
Even though there are add-ons for almost any type of software, it is mostly the browser add-ons which have undergone critical security scrutiny. For example, Google Chrome has an add-on (or \textit{browser extension}) ecosystem, where the add-ons themselves~\cite{jagpal2015trends}, their architecture~\cite{carlini2012evaluation} and protection mechanisms~\cite{heule2015most} have been undergone security evaluation. Similar vetting has been done for Firefox add-ons~\cite{bandhakavi2010vex,barua2013protecting,barth2010protecting}. As more applications are moving to the cloud, we believe that cloud-application add-ons deserve the same attention from the security research community as browser add-ons.

Text editors also have add-on ecosystems (e.g.~Sublime plugins) that have been recently criticized for security vulnerabilities. Azouri Dor analyzed several text editors, and found that it is possible for a malicious add-on to achieve privilege escalation on the victim's computer~\cite{texteditorplugins}. The attack vector here involves crafting a malicious add-on and tricking the victim to install and use it within the text editor. Our attack vector, on the other hand, simply involves injecting a malicious script in a document or other item to be shared with the victim.

\section{Conclusion}
\label{sec:conclusion}

Add-ons in cloud applications are a relatively new phenomenon, whose vulnerabilities have not been widely studied. In this work, we analyzed the security of these add-ons, and we found that flaws in add-ons may introduce new security threats to their host applications. In particular, the add-ons do not always take care when processing untrusted input, which can make them vulnerable to XSS attacks. The attacker can inject malicious scripts into shared documents or emails, which are then processed by the vulnerable add-on. Our study demonstrated that such vulnerable add-ons appear in the wild and that exploiting them is not difficult. Moreover, it seems that cloud-application vendors could do more to limit what the attacker can do once its XSS code is running in the add-on.

{\small
\bibliographystyle{plain}
\bibliography{references}
}


\end{document}